# Towards Climate Neutrality: A Comprehensive Overview of Sustainable Operations Management, Optimization, and Wastewater Treatment Strategies

Vasileios Alevizos [1], Ilias Georgousis[2] and Anna-Maria Kapodistria[3]

[1] Karolinska Institue
[2] International Hellenic University
[3] Linnaeus University

**Abstract:** Various studies have been conducted in the fields of sustainable operations management, optimization, and wastewater treatment, yielding unsubstantiated recovery. In the context of Europe's climate neutrality vision, this paper reviews effective decarbonization strategies and proposes sustainable approaches to mitigate carbonization in various sectors such as building, energy, industry, and transportation. The study also explores the role of digitalization in decarbonization and reviews decarbonization policies that can direct governments' action towards a climate-neutral society. The paper also presents a review of optimization approaches applied in the fields of science and technology, incorporating modern optimization techniques based on various peer-reviewed published research papers. It emphasizes non-conventional energy and distributed power generating systems along with the deregulated and regulated environment. Additionally, this paper critically reviews the performance and capability of micellar enhanced ultrafiltration (MEUF) process in the treatment of dye wastewater. The review presents evidence of simultaneous removal of co-existing pollutants and explores the feasibility and efficiency of biosurfactant instead of chemical surfactant. Lastly, the paper proposes a novel firm-regulator-consumer interaction framework to study operations decisions and interactive cooperation considering the interactions among three agents through a comprehensive literature review on sustainable operations management. The framework provides support for exploring future research opportunities.

**Keywords:** Sustainable operations management, optimization, wastewater treatment, decarbonization strategies, carbonization, building, energy, industry, digitalization, decarbonization policies.

## 1. Introduction

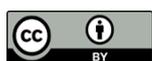

Emphasis is placed on non-conventional energy and distributed power generating systems, as well as deregulated and regulated environments. Furthermore, a critical appraisal of micellar enhanced ultrafiltration (MEUF) process performance and capability in dye wastewater treatment is conducted. Evidence of simultaneous removal of co-existing pollutants is presented, and an investigation into the feasibility and efficiency of biosurfactant in lieu of chemical surfactant is explored [1]. Finally, a novel firm-regulator-consumer interaction framework is proposed for studying operations decisions and interactive cooperation, considering interactions among three agents through an exhaustive literature review on sustainable operations management. This framework offers a foundation for delving into future research opportunities.

*Background and Motivation*

Climate

Spanning back to the Minoan civilization around 3,200 BC, Greece's wastewater management history is enriched with the initial development of drainage and sewerage systems, along with other sanitary infrastructures. Progressively, these technological advancements were adopted on the Greek mainland during subsequent eras of Greek civilization, encapsulating the Mycenaean, Classical, Hellenistic, and Roman periods [2].

*Objectives and Scope of the Study*

Research on achieving climate neutrality and greenhouse gas emissions reduction incorporates investigation into governance, economic tools, technology, dialogue, to promote sustainable development and mitigate climate change. Analyzed are potential of non-conventional vehicles like battery electric, compressed natural gas, hydrogen fuel cell electric vehicles for greenhouse gas emissions reduction from land transportation [3,4]. Control of greenhouse gas emissions might be significantly influenced by economic instruments such as carbon pricing, carbon tax [5], with green finance and circular economy providing support for transition towards carbon neutrality. Gained understanding of climate change economics' principles and methodologies provide crucial insights [6]. Lastly, witnessed are advancements in wastewater infrastructure construction in Greece, yet challenges persist [7,8].

*Structure of the Paper*

Initially, an introduction is provided, whereby context and background information are elucidated, and objectives and hypotheses are outlined. Subsequently, a chapter on achieving climate neutrality in Europe through decarbonization strategies is presented, wherein various approaches and technologies are thoroughly examined, and their effectiveness in reducing greenhouse gas emissions is assessed. A discussion follows, wherein findings are critically analyzed, and potential limitations, implications, and areas for further research are identified [1]. Lastly, a conclusion is drawn, synthesizing key insights gleaned from chapters, and highlighting contributions made to the field of climate change mitigation [2].

## 2. Achieving Climate Neutrality in Europe through Decarbonization Strategies

Climate neutrality in Europe has been envisioned as a critical objective, with decarbonization strategies being extensively pursued to achieve this goal. Simultaneously, innovative wastewater treatment methods, such as Micellar Enhanced Ultrafiltration (MEUF), have been employed to address environmental concerns. Through the amalgamation of Europe's climate neutrality vision, decarbonization efforts, and the implementation of cutting-edge technologies like MEUF, a sustainable future is being forged for the region. By intertwining these essential topics, a comprehensive understanding of the multifaceted approach to environmental preservation and climate action can be gleaned [1].

*Advancing Towards Europe's Climate Neutrality Ambitions*

Progress towards Europe's climate neutrality ambitions has been steadily advanced through the implementation of comprehensive decarbonization strategies. The adoption of innovative technologies, such as Micellar Enhanced Ultrafiltration (MEUF), has been promoted to ensure effective wastewater treatment and to address environmental concerns [9]. A sustainable future for the region has been facilitated by the harmonious integration of these efforts, thereby enabling a multi-faceted approach to environmental preservation and climate action. Consequently, a deeper understanding of the intricate interplay between climate neutrality, decarbonization, and cutting-edge technologies has been provided, highlighting the commitment of European nations to achieving a sustainable and environmentally responsible future [1].

*Decarbonization Strategies in Key Sectors*

Decarbonization strategies in key sectors have been increasingly prioritized as a crucial component of Europe's climate neutrality efforts. In the energy sector, a significant transition to renewable sources has been witnessed, while the reliance on fossil fuels has been gradually diminished. The transportation industry has been revolutionized by the widespread adoption of electric vehicles, and improvements in public transport systems have been made to reduce the carbon footprint. Industrial processes have been reevaluated and optimized to minimize greenhouse gas emissions and sustainable practices have been integrated into agriculture and land use. As a result, the transition towards a low-carbon economy has been accelerated, and the ambition of achieving climate neutrality in Europe has been brought closer to realization [10].

2.1.1. Building Sector

Significant strides have been made in the building sector as part of the concerted efforts towards achieving climate neutrality in Europe. Energy-efficient practices and the use of eco-friendly materials have been increasingly prioritized, leading to the construction of greener, more sustainable buildings [11]. The adoption of passive house designs and the incorporation of renewable energy sources, such as solar panels and geothermal heating systems, have been embraced across the region. Furthermore, retrofitting older buildings with improved insulation and energy-efficient technologies has been widely implemented to reduce the overall energy consumption and greenhouse gas emissions attributable to the built environment [12].

In addition to these technological advancements, policy measures have been enacted to facilitate the transition towards a more sustainable building sector. Stringent energy performance standards have been established by European governments, with financial incentives and support programs being offered to encourage the construction of energy-efficient buildings and the renovation of existing structures. Collaborative efforts among architects, engineers, urban planners, and policymakers have been fostered to develop innovative solutions that address the unique challenges posed by urbanization and climate change. By focusing on the building sector as a key component of the decarbonization strategy, Europe continues to demonstrate its commitment to creating a sustainable and climate-resilient future [13].

2.1.2. Energy Sector

Identified as a crucial component for pursuing climate neutrality in Europe, the energy sector's significant greenhouse gas emissions are attributed to energy production and consumption. Exploration and implementation of diverse low-carbon, renewable energy sources across Europe, for replacing fossil fuels, resulted in a considerable carbon footprint reduction [14]. Intensified efforts for energy efficiency enhancement and conservation promotion have centered on energy consumption optimization in various sectors. Innovative technologies for improved renewable energy integration into grids, investment encouragement in research and development, and promotion of breakthrough energy technologies contribute to resilience, sustainability, economic growth, and job creation, underlining a green economy's potential [15].

2.1.3. Industrial Sector

The industrial sector has been recognized as a key contributor to greenhouse gas emissions, and as such, it is imperative that significant decarbonization efforts are focused on this domain. Various measures have been adopted to reduce the environmental impact of industrial processes, including optimizing energy consumption, using renewable energy sources, and implementing innovative waste management techniques. Additionally, the adoption of circular economy principles has been encouraged, in which resource efficiency and waste reduction are prioritized, thereby promoting sustainable industrial practices [16].

In recent years, the application of advanced technologies, such as Micellar Enhanced Ultrafiltration (MEUF), has been widely embraced in the industrial sector to address wastewater treatment challenges. By utilizing these cutting-edge methods, industries have been able to reduce their environmental footprint and support Europe's climate neutrality ambitions. Furthermore, the collaboration between the public and private sectors has been instrumental in driving research and development in environmentally friendly technologies, leading to the creation of innovative solutions that support a cleaner and more sustainable industrial sector. As Europe moves towards achieving its climate goals, the continued commitment to decarbonization within the industrial sector remains a crucial component of the overall strategy.

*Digitalization as a Catalyst for Decarbonization Efforts*

Digitalization's role as a catalyst for decarbonization efforts across various sectors in Europe is increasingly recognized [17]. Advanced technologies like artificial intelligence, big data, IoT have resulted in substantial improvements in energy efficiency, resource management, and greenhouse gas emissions reduction. Enhanced monitoring and control over energy consumption patterns have been facilitated, processes optimized. A key enabler of Europe's low-carbon economy transition, digitalization accelerates the transition from fossil fuels to renewables, contributes to reducing Europe's carbon footprint, and encourages the adoption of electric vehicles through smart charging infrastructure. Industrial transformation through digitalization results in more sustainable practices, promotes the circular economy, minimizes energy consumption and waste, aligning with Europe's decarbonization and climate neutrality goals. Digitalization fosters sustainable development and accelerates Europe's transition towards a low-carbon future [17].

**3. Discussion**

Compliance with the Urban Wastewater Treatment Directive (UWWTD) and incorporation of water reuse into water resource management strategies are being pursued by Athens Water Supply & Sewerage Company (EYDAP S.A.) through several initiatives in Greece. EU Cohesion funds co-financing has been authorized for two significant wastewater projects in East Attica, aimed at producing treated effluent wastewater suitable for limitless irrigation and urban reuse. Additionally, another wastewater plan is being developed to produce reclaimed water for aquifer recharge, while public datasets related to water supply and wastewater management are being utilized for enhancing the efficiency of these initiatives. Greece is following the Swedish paradigm in wastewater management, learning from key success factors such as a holistic policy approach, integration of recycling and energy recovery, and use of economic instruments to incentivize positive practices, all while considering the country's specific needs and alignment with European Union policies and international technological trends.

*3.1. Wastewater Management in Greece*

To comply with the Urban Wastewater Treatment Directive (UWWTD) and include water reuse in its water resources management strategy, the Athens Water Supply & Sewerage Company (EYDAP S.A.) is working on several wastewater management initiatives. Two significant wastewater projects in East Attica (Rafina/Artemida and Marathon agglomerations) have been authorized for EU Cohesion funds co-financing and implementation by EYDAP S.A [18]. The goal of these programs is to create treated effluent wastewater that meets national criteria for limitless irrigation and urban reuse. Another wastewater plan is being developed incorporating the agglomerations of Koropi and Paiania, which will produce reclaimed water appropriate for aquifer recharge to restore the water quality of groundwater bodies [19].

In addition, a plethora of public data sets related to water supply and wastewater management, such as computer modeling of water supply and sewerage networks, the implementation of an integrated SCADA system, and the history of water supply and wastewater management in Paris and the Republic of Belarus, could be used to further improve the efficiency of such initiatives [19].

*Wastewater Management in Sweden*

In 2013, the Swedish Environmental Agency recommended a national aim for increasing phosphorus recycling from wastewater sludge. Sweden has more than 80 years of experience protecting water quality, and the creation of phosphorus removal technology may be a Swedish contribution to advanced knowledge. Source separation systems have been found to be an efficient method of recovering nutrients and energy from wastewaters in both rural and urban settings, with research on the nutrient recovery potential and life cycle consequences of source separation systems undertaken in northern Finland and Sweden. In Sweden, exploratory research looked at how local administration and municipally held enterprises influence the governance of industrial symbiosis in the water and sewage sectors. Finally, a special issue was published on municipal wastewater management in 2021 [20].

*3.2. Greek Government following Swedish paradim in waterwast management*

Several key success factors from Sweden's waste management paradigm can be learned and applied by Greece to enhance its waste management practices. A holistic policy approach is employed by Sweden, which addresses diverse public demands by integrating waste management with other environmental and economic policies. Recycling and energy recovery are integrated in Sweden, as 99% of municipal solid waste is recycled and energy is harnessed, with less than 1% going to landfills [21]. Taxes and tariffs are utilized by Sweden as economic instruments to discourage harmful practices and incentivize positive ones, such as recycling and energy recovery. Autonomy is given to Swedish municipalities, allowing them to have the economic and operational capacity to manage waste collection and treatment systems. By adopting these key success factors, improvements can be made to Greece's waste management practices, reductions in environmental pollution can be achieved, and sustainable development can be promoted. Careful consideration should be given to Greece's specific needs, such as interactions with the extensive tourism sector, and alignment of waste management strategies with the European Union's framework policies and international technological trends. In doing so, wastewater treatment plants can be transformed into sites where energy is efficiently used or produced, resources are recovered and reused, and environmental sustainability is practiced overall [21].

## 4. Conclusions

In compliance with the Urban Wastewater Treatment Directive (UWWTD), several wastewater management initiatives are being undertaken by the Athens Water Supply & Sewerage Company (EYDAP S.A.) to incorporate water reuse into Greece's water resources management strategy. Authorization for EU Cohesion funds co-financing has been granted for two significant wastewater projects in East Attica (Rafina/Artemida and Marathon agglomerations), which are being implemented by EYDAP S.A [18]. National criteria for limitless irrigation and urban reuse are aimed to be met by the creation of treated effluent wastewater through these programs. Another wastewater plan is being developed, which will incorporate the agglomerations of Koropi and Paiania, with the production of reclaimed water appropriate for aquifer recharge to restore the water quality of groundwater bodies [19].

Furthermore, the efficiency of such initiatives can be improved by utilizing a plethora of public data sets related to water supply and wastewater management, such as computer modeling of water supply and sewerage networks, the implementation of an integrated SCADA system, and the history of water supply and wastewater management in Paris and the Republic of Belarus [19].